\documentstyle[aps,prb,psfig]{revtex}
\begin{document}

\title{Thermal conductivity in the Vortex State of the
Superconductor UPd$_2$Al$_3$}
\author{L. Tewordt and D. Fay}
\address{I. Institut f\"ur Theoretische Physik,
Universit\"at Hamburg, Jungiusstr. 9, 20355 Hamburg, 
Germany}
\date{\today}
\maketitle
\begin{abstract}
     The magneto-thermal conductivity $\kappa$ is calculated for the vortex
state of UPd$_2$Al$_3$ by assuming horizontal gap nodes. The Green's
function method we employed takes into account the effects of supercurrent
flow and Andreev scattering on the quasiparticles due to Abrikosov's vortex
lattice order parameter. The calculated angular dependence of $\kappa_{yy}$
for field rotation $\theta_0$ in the ac-plane depends strongly on field strength
$H\,$, impurity scattering, anisotropy of the Fermi velocity, and temperature.
For finite temperatures and the clean unitary scattering limit we get 
qualitative agreement with recent experiments for all four proposed gap
functions having horizontal line nodes at $c\,k_z = 0\,$, $\pm\pi/4\,$, and
$\pm\pi/2\,$.
\end{abstract}
\vspace{0.5in}
       Angle-dependent magneto-thermal conductivity is a powerful tool for
determining the nodal structure of the gap in unconventional
superconductors. Recently, the thermal conductivity has been measured
in the heavy-Fermion superconductor UPd$_2$Al$_3$ for a variety of 
magnetic field orientations. \cite{Watanabe} The thermal conductivity
$\kappa_{yy}$ displays two-fold oscillations when the magnetic field
$\mbox{\boldmath$H$}$ is rotated in the ac-plane, while no oscillations are
observed when $\mbox{\boldmath$H$}$ is rotated within the basal 
ab-plane. These results provide strong evidence that the gap function
$\Delta(\mbox{\boldmath$k$})$ has horizontal line nodes orthogonal to the
c-axis. Four gap models have been proposed and denoted in 
Ref.~\onlinecite{Watanabe} as types I - IV, respectively: 
$\Delta(\mbox{\boldmath$k$}) \propto \sin\chi\,$, $\cos\chi\,$, $\sin2\chi\,$, 
and $\cos2\chi\,$ (with $\chi = ck_z$). Cooper pairing mediated by magnetic
excitons yields the highest $T_c$ for the model $\cos\chi\,$. \cite{McHale}
The magneto-thermal conductivity for these models has been calculated by
the method of the Doppler shift of the energy of quasiparticles due to the
circulating supercurrent flow of the vortices. \cite{Won,Thalmeier} This effect
depends sensitively on the angle between $\mbox{\boldmath$H$}$ and the
direction of the nodes of the gap. Comparison of these Doppler shift results
with the data indicates that the model type IV, 
$\Delta(\mbox{\boldmath$k$})=\Delta\cos2\chi\,$, gives the most consistent 
description of the experiments of Ref.~\onlinecite{Watanabe}. \cite{Won}
A new Doppler shift approximation however yields the opposite result, i.e.,
that the models of types I - III are in better agreement with experiment.
\cite{Thalmeier}

    The purpose of the present paper is to calculate the magneto-thermal
conductivity for horizontal gap nodes by another method which takes into
account, beside the effect of the supercurrent flow, the Andreev scattering 
off the vortex cores. \cite{BPT} These effects are calculated from the real
and imaginary parts of the Andreev scattering self energy for the 
quasiparticle Green's function in the presence of Abrikosov's vortex lattice
order parameter. This method was applied to Sr$_2$RuO$_4$ by assuming
vertical and horizontal nodes of the superconducting gap. \cite{TF1} An
equivalent method based on the quasiclassical equations and linear
response theory \cite{Pesch} provides much more compact expressions for
the density of states and thermal conductivity. \cite{Vek} We have shown that
the latter expressions yield very nearly the same results as the original
expressions derived in Refs.~\onlinecite{BPT} and \onlinecite{TF1}
(see Ref.~\onlinecite{TF2}).

     The Pesch-approximation \cite{Pesch} yields for the spatial average of
the normalized density of states:
\begin{equation}
N(\omega,\mbox{\boldmath$k$}) = \mbox{Re}\,
g(\omega,\mbox{\boldmath$k$})\:;   
\quad g(\omega,\mbox{\boldmath$k$}) = \left\{
1+8|\Delta(\mbox{\boldmath$k$}) |^2 \,
[\Lambda/v_{\perp}(\mbox{\boldmath$k$})]^2 \,
 [1+i\sqrt{\pi}\,z\, w(z)]\right\}^{-1/2}
\label{N}
\end{equation}
where
\begin{equation}
z=2[\omega+i\,\Sigma_{\mbox{i}}(\omega)]
[\Lambda/v_{\perp}(\mbox{\boldmath$k$})]\:;
\qquad \Lambda=(2eH)^{-1/2}\,.
\label{z}
\end{equation}
Here $v_{\perp}(\mbox{\boldmath$k$})$ is the component of the Fermi 
velocity perpendicular to the magnetic field $\mbox{\boldmath$H$}\,$, and
$\Sigma_{\mbox{i}}(\omega)$ is the self energy for impurity scattering
which is calculated self-consistently in the t-matrix approximation. 
\cite{TF2} The integrand of the $\omega$-integral for the thermal 
conductivity is proportional to 
$\left(\frac{\omega}{T}\right)^2\cosh^{-2}\left(\frac{\omega}{2T}\right)
\mbox{Re}\,g(\omega,\mbox{\boldmath$k$})$ times the transport
scattering time $\tau(\omega,\mbox{\boldmath$k$})$ where\, \cite{Vek}
\begin{equation}
\frac{1}{2\,\tau(\omega,\phi)} = \mbox{Re}\,\Sigma_{\mbox{i}}(\omega)
+2\sqrt{\pi}\,[\Lambda/v_{\perp}(\mbox{\boldmath$k$})]
|\Delta(\mbox{\boldmath$k$}) |^2
\frac{\mbox{Re}[g(\omega,\mbox{\boldmath$k$})w(z)]}
{\mbox{Re}\,g(\omega,\mbox{\boldmath$k$})}\,.
\label{Tau}
\end{equation}
The first term in Eq.[\ref{Tau}] is the scattering rate due to impurities which
reduces in the normal state to $\Gamma=1/2\tau_n\,$. The second term is
the Andreev scattering rate due to the vortex cores.

     For the heavy-Fermion superconductor UPd$_2$Al$_3$ we consider
only the cylindrical Fermi surface and approximate the Fermi velocity by
\begin{equation}
\mbox{\boldmath$v$}_F=v_a\,[\cos\phi\:\mbox{\boldmath$e$}_a + 
\sin\phi\:\mbox{\boldmath$e$}_b + 
\varepsilon\,\sin2\chi\:\mbox{\boldmath$e$}_c]\,,\quad
(\chi=ck_z\,,\:\, -\pi/2<\chi<+\pi/2)\,,
\label{v}
\end{equation}
where $\varepsilon=v_c/v_a\,$. Furthermore we assume the following gap
function with horizontal nodes:
\begin{equation}
\Delta(\mbox{\boldmath$k$})=\Delta\cos\chi\,;\quad(\chi=ck_z)\,;
\quad\Delta=\Delta_0[1-h^2]^{1/2}\,;\quad(h=H/H_{c2})\,.
\label{Delta}
\end{equation}
For a field direction given by polar and azimutal angles $\theta_0$ and
$\phi_0\,$, i.e.,
\begin{equation}
\mbox{\boldmath$H$}=H\,[\sin\theta_0\cos\phi_0\:\mbox{\boldmath$e$}_a +
\sin\theta_0\sin\phi_0\:\mbox{\boldmath$e$}_b + 
\cos\theta_0\:\mbox{\boldmath$e$}_c]\, ,
\label{H}
\end{equation}
the component of $\mbox{\boldmath$v$}_F$ perpendicular to 
$\mbox{\boldmath$H$}$ becomes:
\begin{equation}
v_{\perp}(\mbox{\boldmath$k$})=v_F\left\{1-\frac{[\sin\theta_0\cos(\phi-\phi_0)
+ \varepsilon\cos\theta_0\sin2\chi]^2}
{[1+\varepsilon^2\,\sin^22\chi]}\right\}^{1/2}\, .
\label{vperp}
\end{equation}
From this expression it is obvious that the combination
$\Delta(\mbox{\boldmath$k$})/v_{\perp}(\mbox{\boldmath$k$})$ occuring
in Eqs. (\ref{N}) and (\ref{Tau}) depends sensitively on the direction of
$\mbox{\boldmath$H$}$ relative to $\mbox{\boldmath$k$}$ and thus relative
to the gap nodes. For this model we have calculated $\kappa_{yy}/\kappa_n$
for field orientations $\mbox{\boldmath$H$}$ in the xz-plane
($\,\phi_0=0\,$; $\;0\le\theta_0\le\pi\,$). For the anisotropy parameter 
$\varepsilon=v_c/v_a$ in Eq.(\ref{v}) we take the value $\varepsilon=0.6\,$.
We first consider an impurity scattering rate 
$\delta=\Gamma/\Delta_0=0.024$ and a scattering phase shift
$\rho=0.9(\pi/2)\,$. In Fig. 1a we show $\kappa_{yy}/\kappa_n$ versus the 
reduced field $h$ for $\theta_0 = 0\,$ (solid curve) and $\theta_0 = \pi/2\,$
(dashed curve) for $T=0\,$ ($\omega=0$\,). We have also plotted the 
relative amplitude of the oscillation, 
$\Delta\kappa= [\kappa_{yy}(\theta_0=0)-\kappa_{yy}(\theta_0=\pi/2)]/
\kappa_{yy}(\theta_0=0)\,$. In Fig. 1b we show two examples of the 
oscillation as functions of the polar angle $\theta_0$ for the fixed field
strengths $h=0.05$ and $h=0.3\,$. Our results for the unitary limit (phase
shift $\rho=\pi/2\,$) are nearly the same as those shown in Fig. 1. Different
results are obtained in Ref.~\onlinecite{Thalmeier} for phase shift
$\rho=0.9(\pi/2)\,$ where a small region (denoted by I in 
Ref.~\onlinecite{Watanabe}) exists in which $\kappa_{yy}$ decreases with 
increasing $h$ and the amplitude of the oscillation $\Delta\kappa$ becomes
negative. This would be in agreement with the data of 
Ref.~\onlinecite{Watanabe}.

     The measured angular variation of $\kappa_{yy}(\theta_0)$ is decomposed
into a constant and a term with twofold symmetry with respect to the 
$\theta_0$ rotation, i.e., $C_{yy}\cos2\theta_0\,$. \cite{Watanabe} In the
region denoted by II in Ref.~\onlinecite{Watanabe}, where $\kappa_{yy}$
depends linearly on $H\,$, $C_{yy}$ is positive. In the small region I near
$H=0\,$, where $\kappa_{yy}$ decreases with increasing $H\,$, the 
measured amplitude $C_{yy}$ becomes negative.

     We have not been able to confirm the results of Ref.~\onlinecite{Thalmeier}
where the origin of the sign change of $\Delta\kappa$ at low fields is the
tiny deviation of the impurity scattering phase shift $0.9(\pi/2)$ from the
unitary limit $\pi/2\,$. We show now that this sign change is actually due
to the finite temperature. We estimate the finite temperature effect by
considering finite values of $\Omega=\omega/\Delta_0$ where $\omega$
occurs in Eqs. (\ref{z}) and (\ref{Tau}) and is the integration variable for the
$\omega$-integral yielding $\kappa\,$. Since this integrand is strongly
peaked at $\omega/T\simeq 2.4\,$, the temperature corresponding to
$\Omega$ is approximately given by 
$\Omega\simeq 2.4(T/\Delta_0)\sim T/T_c\,$. We have verified that this is
a good approximation by numerical computation of the $\omega$-integral
at finite $T\,$. As an example,  in Fig. 2a we show $\kappa_{yy}/\kappa_n$
versus $h$ for $\Omega=0.3$ for the impurity scattering rate $\delta=0.024$
and phase shift $\pi/2\,$. The solid curve refers to the angle $\theta_0=0$
and the dashed curve to $\theta_0=\pi/2\,$. One observes now a
small range from $h=0$ to about $h=0.1$  where the amplitude 
$\Delta\kappa$ of the oscillation is negative. In the enlarged scale of
Fig. 2b it is seen more clearly that in this region $\kappa_{yy}/\kappa_n$
bends upwards for decreasing $h\,$, which also occurs in region I in the 
experiments. In Fig. 2c we have plotted the oscillation of $\kappa_{yy}$ as
a function of the polar angle $\theta_0$ for fixed field strengths
$h=0.05\,$, 0.3, and =0.5. One sees that oscillations of approximately 
twofold symmetry in $\theta_0$ occur.  In the region of approximately
linear field dependence of $\kappa_{yy}$ (see Fig. 2a), which corresponds
to region II in the experiments, the amplitude $\Delta\kappa$ corresponding
to $C_{yy}$ in Ref.~\onlinecite{Watanabe} is positive. In the region of low
fields, corresponding to region I of Ref.~\onlinecite{Watanabe}, 
$\Delta\kappa$ is in fact negative for this finite temperature. It is important to
note that the experiments are carried out at a finite temperature of about 
0.4 K. \cite{Watanabe}

      We discuss now the dependence of our results on the various
parameters. First we consider the effect of impurity scattering. For a larger
scattering rate, for example, $\delta=\Gamma/\Delta_0=0.1\,$, and the
unitary limit, we obtain for $\Omega=0.3$ the functions $\kappa_{yy}(h)$
and $\Delta\kappa$ shown in Fig. 3a. Comparison with the results shown in
Fig. 2a for $\delta=0.024\,$, and all other parameters the same, shows
that the slope of $\kappa_{yy}$ in the linear range of region II is larger, 
while it is smaller in the region, denoted by III in Ref.~\onlinecite{Watanabe},
where $\kappa_{yy}/\kappa_n$ rises steeply to one. This field dependence
in Fig. 3a is in better agreement with experiment. However, no region I
occurs where $\kappa_{yy}$ decreases with $h$ and $\Delta\kappa$ 
becomes negative. The function $\kappa_{yy}(\theta_0)$ exhibits an
approximate twofold symmetry in $\theta_0$ for all values of $h\,$. 
For $\delta=0.1$ we have to go to higher values of $\Omega$ to obtain a
sign change in $\Delta\kappa$ near $h=0\,$. This is shown in Fig. 3b 
where $\Delta\kappa$ and $\kappa_{yy}$ are plotted vs $h$ in an enlarged
scale for $\Omega=$ 0.3, 0.4, and 0.5. One sees that for $\Omega=0.4\,(0.5)\,$,
$\Delta\kappa$ becomes negative below $h\simeq0.06\,(0.18)\,$. These
regions of negative $\Delta\kappa$ correspond to region I because 
$\kappa_{yy}$ bends upwards for decreasing $h\,$. Another important 
parameter
is the phase shift for impurity scattering. We have already pointed out that,
contrary to the results of Ref.~\onlinecite{Thalmeier}, we obtain practically
no difference between the results for the unitary phase shift limit
$\pi/2$ and $0.9(\pi/2)\,$. However, for the Born limit, phase shift zero, the 
curve for $\kappa_{yy}(h)$ is much flatter in the lower $h$-range than the
curves for the unitary limit in Figs. 2a and 3. Also,  the amplitude
$\Delta\kappa$ is negative and $|\Delta\kappa|$ is much larger over the
entire range of $h\,$. These results in the Born limit are in agreement with
those obtained in Ref.~\onlinecite{Thalmeier}.

     We come now to the most crucial parameter, $\varepsilon=v_c/v_a$ in
Eq.(\ref{v}), which determines the anisotropy of the Fermi
velocity for the corrugated cylindrical Fermi surface sheet. For the range of
$\varepsilon$-values between about $\varepsilon=0.5$ and 0.7 we obtain
nearly the same results as those shown for $\varepsilon=0.6\,$. These values
agree approximately with the $\varepsilon$ value which is obtained
according to Ref.~\onlinecite{Thalmeier} if we take for $v^2_c$ and $v^2_a$
the average values over the corrugated cylindrical Fermi surface where
$(v_c/v_a)^2 = (1/2)(v_{c0}/v_{a0})^2$ and 
$\alpha=(v_{c0}/v_{a0})^2 \simeq 0.69$ is the appropriate value for 
UPd$_2$Al$_3\,$. In addition to the order parameter $\Delta\propto\cos\chi\,$
(type II), we also obtain, for $\varepsilon$ between 0.5 and 0.7, nearly the
same results for the other order parameters proposed in 
Ref.~\onlinecite{Watanabe}, namely, $\Delta\propto\sin\chi$ (type I),
$\sin2\chi$ (type III), and $\cos2\chi$ (type IV). In all cases $\chi=ck_z\,$.
However, for $\Omega$ small, as $\varepsilon$ increases from 0.7 to about 
0.84, the minimum of $\kappa_{yy}(\theta_0)$ at the field angle 
$\theta_0=(\pi/2)$ changes to a smaller maximum and two minima 
develop at the intermediate angles of about $\theta_0\simeq\pm1.3(\pi/4)\,$. 
It should be pointed out that, for the type IV gap, the minima already start to
form just above $\varepsilon=0.6$ and become more pronounced than for the
type I - III gaps. For increasing $\Omega\,$, the small peak at $\pi/2$ is
reduced and becomes a  flat, broad minimum at about $\Omega=0.3\,$.
Furthermore,  for $\Omega$ above 0.26,the amplitude $\Delta\kappa$ of the 
variation of $\kappa_{yy}(\theta_0)$ becomes negative in a small range 
near $h=0\,$. This behavior is similar to that of $\Delta\kappa$
shown in Fig. 2 for $\varepsilon=0.6$ and $\Omega=0.3$ and again we see 
the importance of finite $\Omega$ (temperature) in reproducing the
experimental results. We remark that, for $\Omega=0.3$ and 
$\varepsilon=0.6\,$, $\Delta\kappa$ is negative below $h=0.05$ for the 
type IV gap while it is negative below $h=0.1$ for the type II gap.

We also find that, for $\Omega$ larger than 0.4, a sign change of 
$\Delta\kappa$ occurs in a small range of $h$ just below $h=1\,$. In this
connection we remark that experiments yield a negative sign of the amplitude
$C_{yy}$ in a region denoted by III in Ref.~\onlinecite{Watanabe} where 
$\kappa_{yy}$ rises steeply with $H\,$. This negative sign has however been
attributed to the anisotropies of the Fermi velocity and the upper critical field 
$H_{c2}\,$. Since we have neglected here the anisotropy of $H_{c2}$ it may
be that our calculated values of $|\Delta\kappa|$ in region III are much 
smaller than the measured values.

    Let us compare our results with those obtained by the Doppler shift
method. For the clean unitary scattering limit and intermediate field
strengths we obtain for the angular variation of $\kappa_{yy}(\theta_0)$ a
positive amplitude $\Delta\kappa= [\kappa_{yy}(\theta_0=0)-
\kappa_{yy}(\theta_0=\pi/2)]/\kappa_{yy}(\theta_0=0)$ for all four of the
gap models of types I - IV proposed in Ref.~\onlinecite{Watanabe}. The first
Doppler shift calculation\, \cite{Won} yielded a positive $\Delta\kappa$ for the 
gaps $\cos2\chi$ and $\sin\chi$ and a negative $\Delta\kappa$ for
$\cos\chi\,$. Another, more recent, Doppler calculation \,\cite{Thalmeier} finds
a positive $\Delta\kappa$ for $\sin\chi\,$, $\cos\chi\,$, and $\sin2\chi$
whereas the variation for the gap $\cos2\chi$ is completely different from 
the form $\cos2\theta_0$ and exhibits pronounced minima at the angles
$\theta_0=\pm 51^o\,$. It is interesting that we obtain similar structure in 
$\kappa_{yy}(\theta_0)$ with minima at intermediate angles 
$\theta_0 \simeq \pm 1.3(\pi/4)$ for all four gap models I - IV if the 
anisotropy parameter $\varepsilon=v_c/v_a$ is increased from 0.6 to 0.84.
It is not surprising that our analytical expressions yield different results
than the Doppler shift method because we take the Andreev scattering of
the quasiparticles by the vortex cores into accoount. It has been shown by
comparison with the exact results of the quasiclassical Eilenberger 
equations (see Fig.7 in Ref. \onlinecite{Dahm}) that the analytical 
expressions \, \cite{BPT,Pesch} provide good approximations over the 
whole field range from $H_{c2}$
to $H_{c1}$ wheras the Doppler shift method is a good approximation only
at low fields. However we want to caution that in the low field regime the
higher Fourier components of the Green's function with respect to the vortex
lattice vectors\, \cite{BPT} might lead to important corrections. Therefore
a fully self-consistent numerical calculation, in particular for the internal field,
would be necessary \cite{Nakai} in order to obtain more reliable results. This
means that the method of Ref. \onlinecite{Nakai} for the density of states 
should  be extended to the calculation of the Andreev scattering from the 
superfluid flow in the single vortex regime which is important for the thermal
conductivity $\kappa\,$. The large effect of this Andreev scattering can be
seen from the fact that, at $T=0$ and impurity scattering phase 
shift = 0.9 ($\pi/2$), $\kappa$ increases for small fields while the Doppler
shift method together with a scattering lifetime due to impurity scattering
only yields an initial drop of $\kappa$ for small fields. \cite{Thalmeier} We
stress again that only at finite temperature ($\Omega$) do we obtain an initial
drop of $\kappa$ together with a sign reversal of the oscillation amplitude 
$\Delta\kappa$ as shown in Figs. 2b and 3b. 

     In summary, in the clean unitary scattering limit, and for not too large
anisotropy $\varepsilon$ of the Fermi velocity, we obtain angular 
variations of $\kappa_{yy}(\theta_0)$ of approximately twofold symmetry
in the field rotation angle $\theta_0$ for all of the four proposed gap models
having horizontal line nodes at $ck_z=\chi=0\,$, $\pm\pi/4\,$, and
$\pm\pi/2\,$. The amplitude $\Delta\kappa$ is positive in the range of field
values $H$ where $\kappa_{yy}$ is nearly linear in $H\,$. At finite 
temperature $\Delta\kappa$ becomes negative at small fields which is in 
agreement with experiment. The range of field values with negative 
$\Delta\kappa$ is smaller for the gap of type IV ($\cos2\chi$) than for
types I - III. Our results depend crucially on the anisotropy parameter
$\varepsilon=v_c/v_a$ for the corrugated cylindrical Fermi surface. For
values of $\varepsilon$ above about 0.7 we find a small maximum at 
$\theta_0 = \pi/2$ and two minima at about $\theta_0\simeq\pm1.3(\pi/4)$
in the curve of $\kappa_{yy}(\theta_0)$ for the gap models I -III. For the 
gap model IV this critical value of $\varepsilon$ is smaller, about 
$\varepsilon =0.6\,$. It should be pointed out that even for a constant gap
the density of states depends strongly on the shape of the Fermi surface
and the field orientation. \cite{Graser} Thus we can say that the angular
variations of $\kappa_{yy}(\theta_0)$ are determined by two competing
effects: the positions of the gap nodes and the amount of anisotropy of the
Fermi velocity given by the shape of the Fermi surface. Here the latter
effect appears to predominate.

     In conclusion, our calculations indicate that at present one cannot
determine which one of the proposed gap models with horizontal line
nodes is most consistent with the measured magneto-thermal
conductivity of UPd$_2$Al$_3\,$. The gap model $\cos\chi$ however 
seems to be somewhat more favorable than $\cos2\chi\,$. Our
calculations also indicate the importance of accounting for finite
temperature and lead us to expect that the value of the field at which the 
amplitude of the oscillation $\Delta\kappa$ changes sign  should decrease
with decreasing temperature. It is thus desirable to have measurements 
available at lower temperatures in very clean samples. It will also be
interesting to see whether additional structure, minima, for example, occur
in the form of $\kappa_{yy}(\theta_0)\,$.
\newpage
\newpage
\begin{figure}
\centerline{\psfig{file=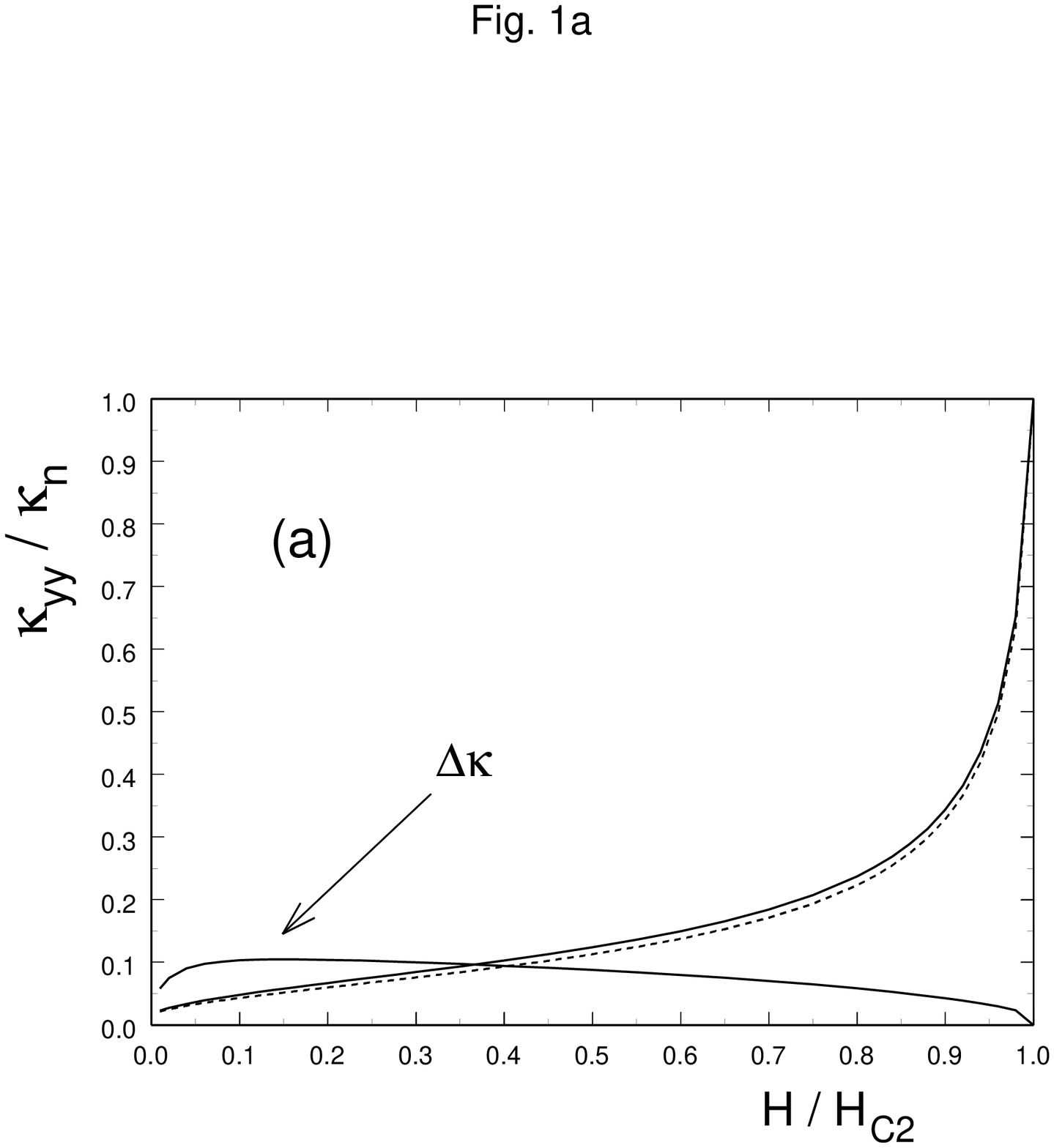,width=18cm,angle=0}}
\vskip -1.5cm
\caption{1a) Thermal conductivity $\kappa_{yy}$ for the gap model I vs reduced 
field $h=H/H_{c2}$ for field direction $\theta_0=0\,$ (solid curve) and 
$\theta_0=\pi/2\,$ (dashed curve). Also shown is the relative amplitude of the
oscillation 
$\Delta\kappa= [\kappa_{yy}(\theta_0=0)-\kappa_{yy}(\theta_0=\pi/2)]/
\kappa_{yy}(\theta_0=0)\,$. The impurity scattering rate is
$\delta=\Gamma/\Delta_0 = 0.024\,$, the phase shift is $0.9(\pi/2)\,$, and
the anisotropy parameter $\varepsilon=v_c/v_a=0.6\,$.}
\label{fig1a}
\end{figure}
\begin{figure}
\centerline{\psfig{file=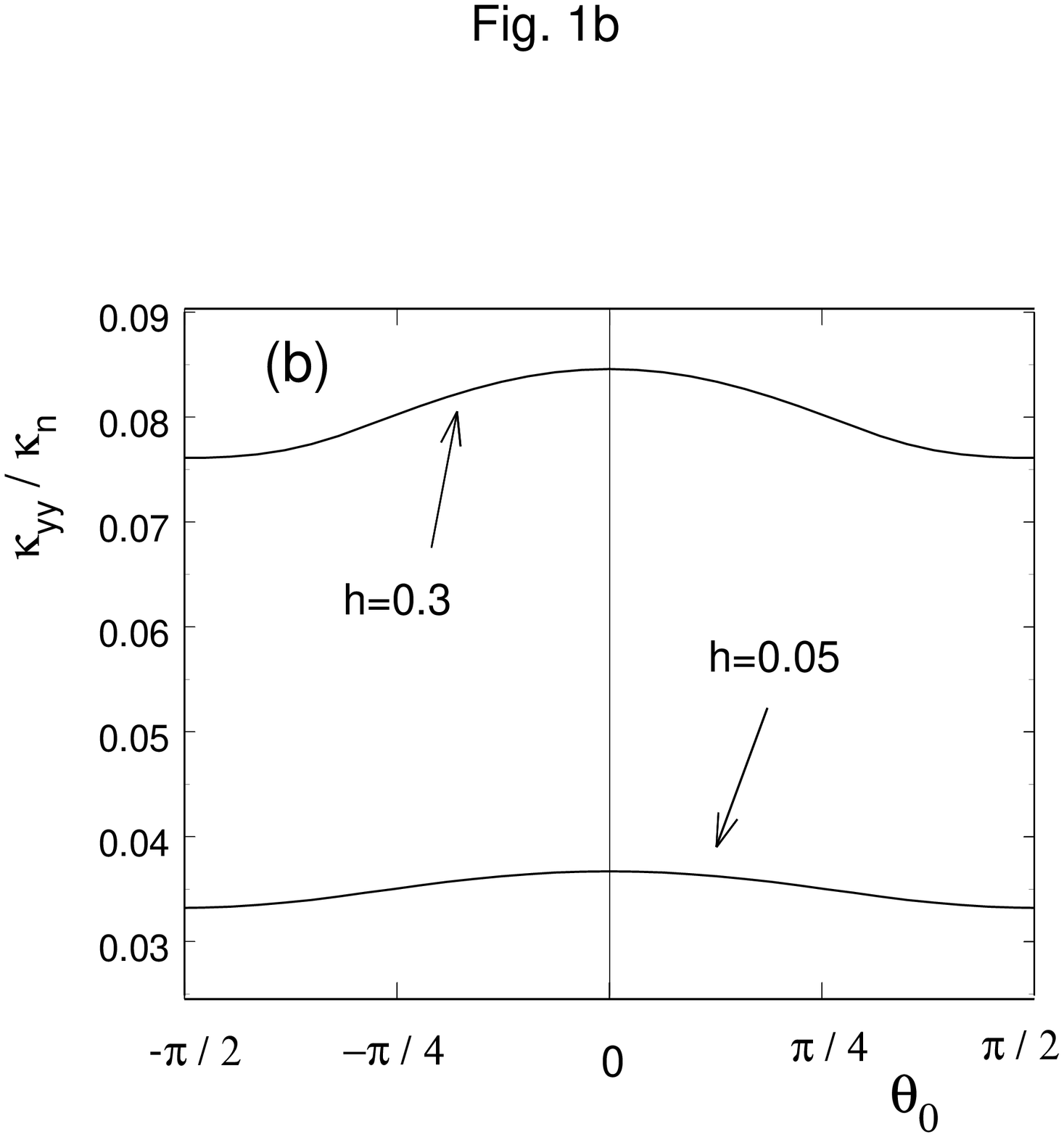,width=18cm,angle=0}}
\vskip -1cm
\caption{1b)  Angular variation of $\kappa_{yy}$ vs polar 
angle $\theta_0$ for field strengths $h=0.05$ and $h=0.3\,$.}
\label{fig1b}
\end{figure}
\begin{figure}
\centerline{\psfig{file=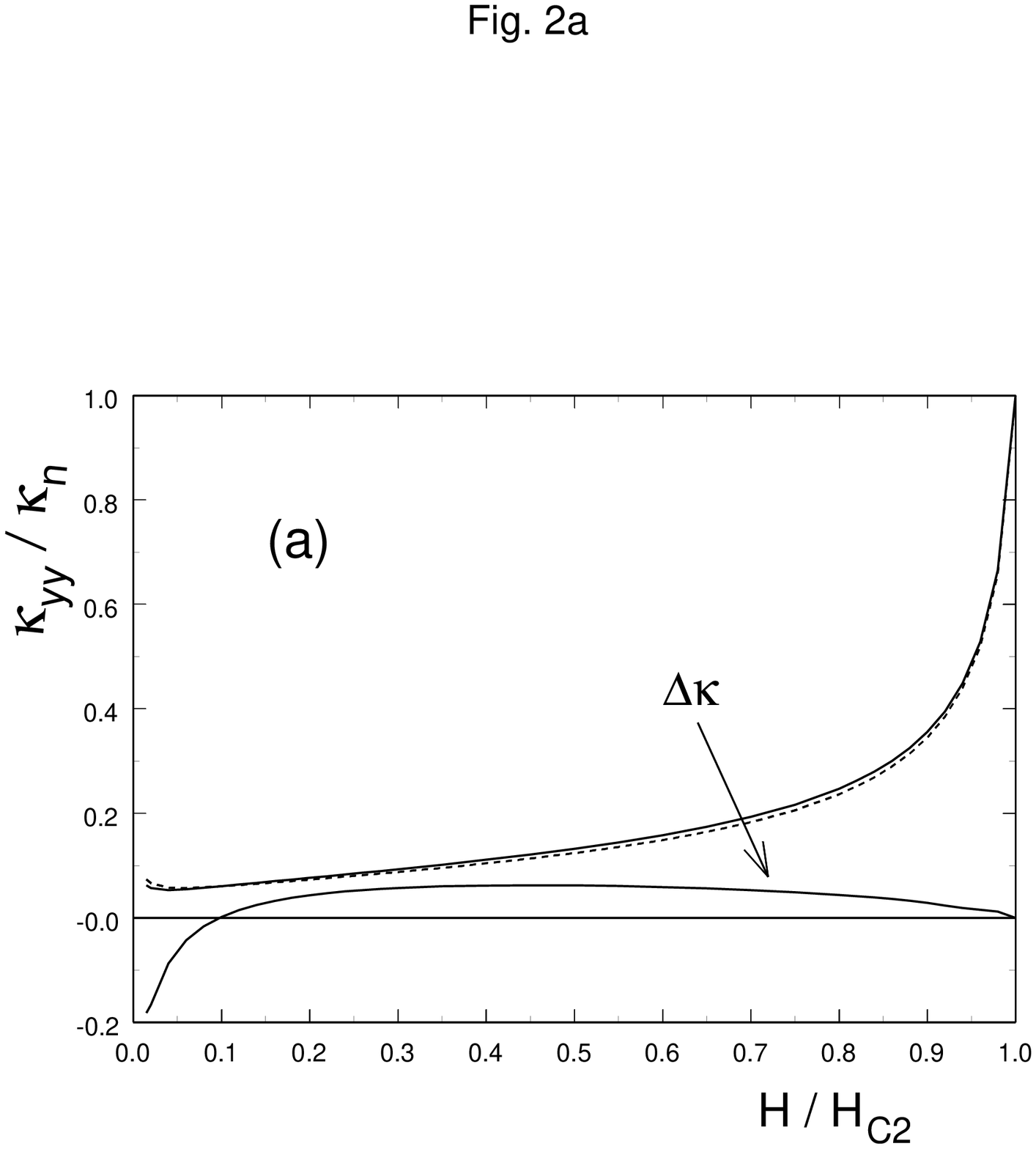,width=18cm,angle=0}}
\vskip -1cm
\caption{2a) $\kappa_{yy}$ vs $h$ for $\theta_0=0\,$ (solid curve) and
$\theta_0=\pi/2\,$ (dashed curve), and $\Delta\kappa$ vs $h\,$, for impurity 
scattering rate $\delta=0.024\,$, phase shift $\pi/2\,$ (unitary limit), and 
$\varepsilon=0.6\,$. The reduced frequency is 
$\Omega=\omega/\Delta_0=0.3$ which corresponds to a finite temperature 
of order $T/T_c\sim\Omega\,$.}
\label{fig2a}
\end{figure}
\begin{figure}
\centerline{\psfig{file=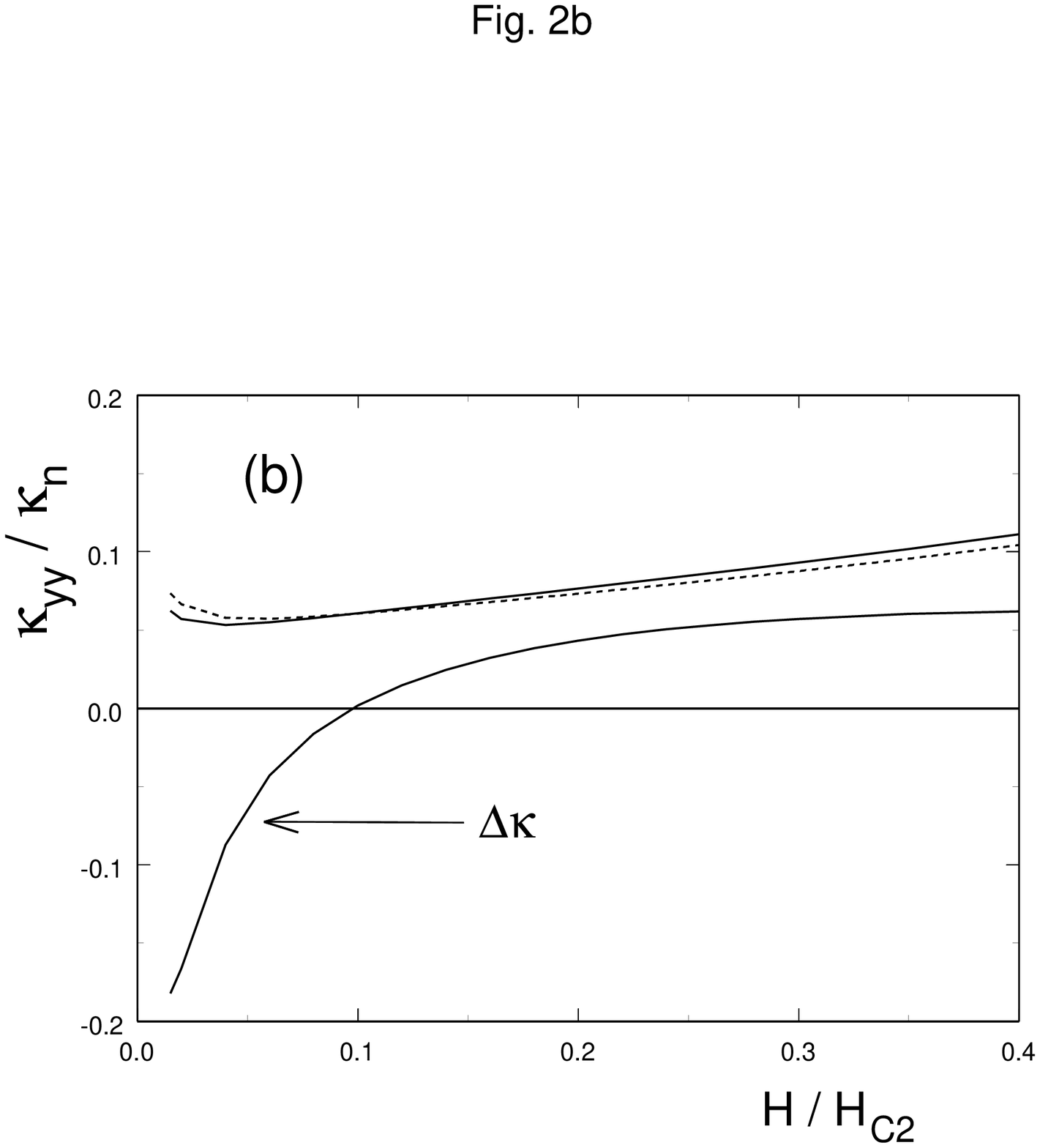,width=18cm,angle=0}}
\vskip -1cm
\caption{2b) the same as in a) for an enlarged scale.}
\label{fig2b}
\end{figure}
\begin{figure}
\centerline{\psfig{file=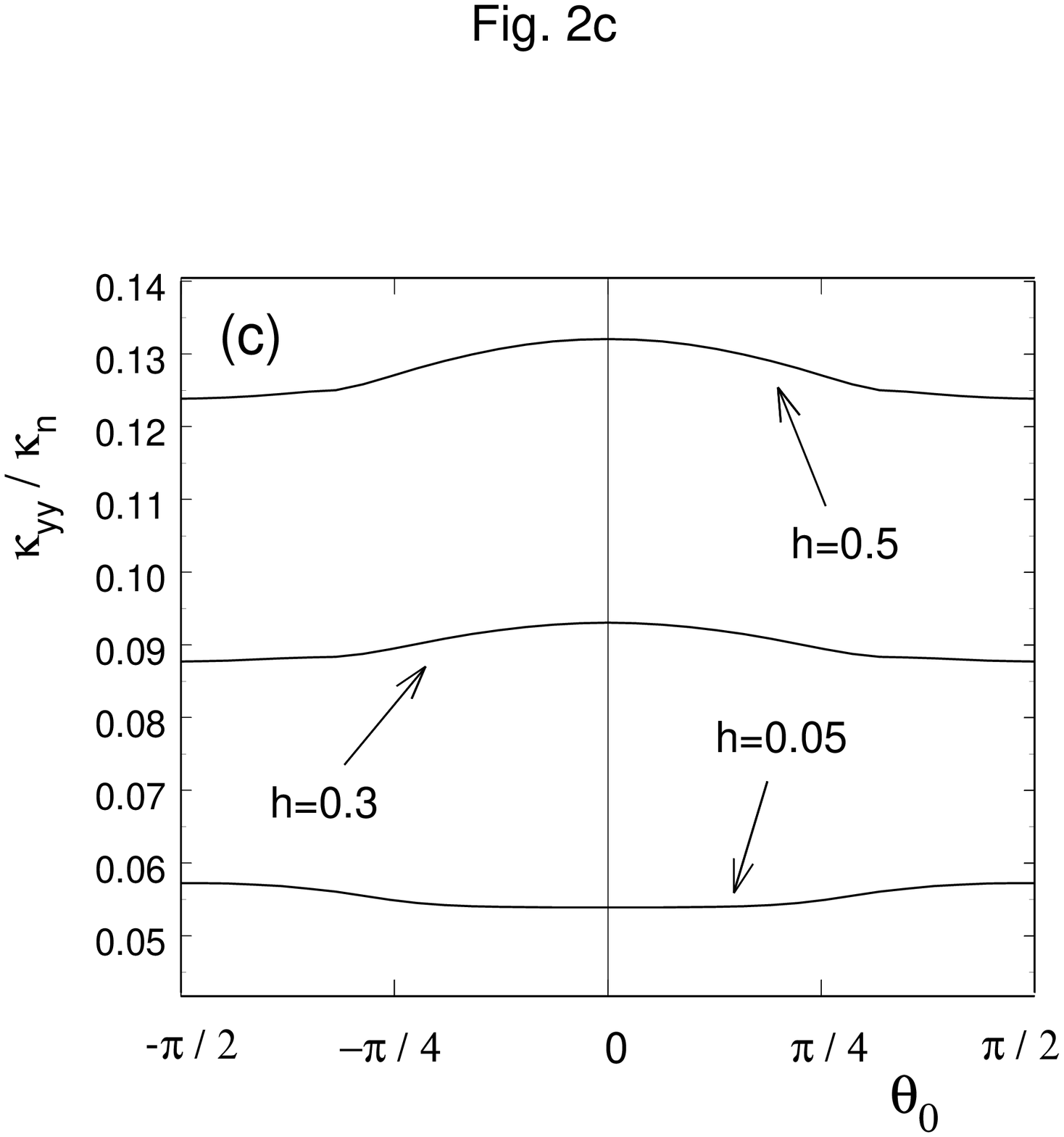,width=18cm,angle=0}}
\vskip -1cm
\caption{2c) Angular variation of $\kappa_{yy}(\theta_0)$ for field strengths 
$h=0.05\,$, 0.3, and 0.5.}
\label{fig2c}
\end{figure}
\begin{figure}
\centerline{\psfig{file=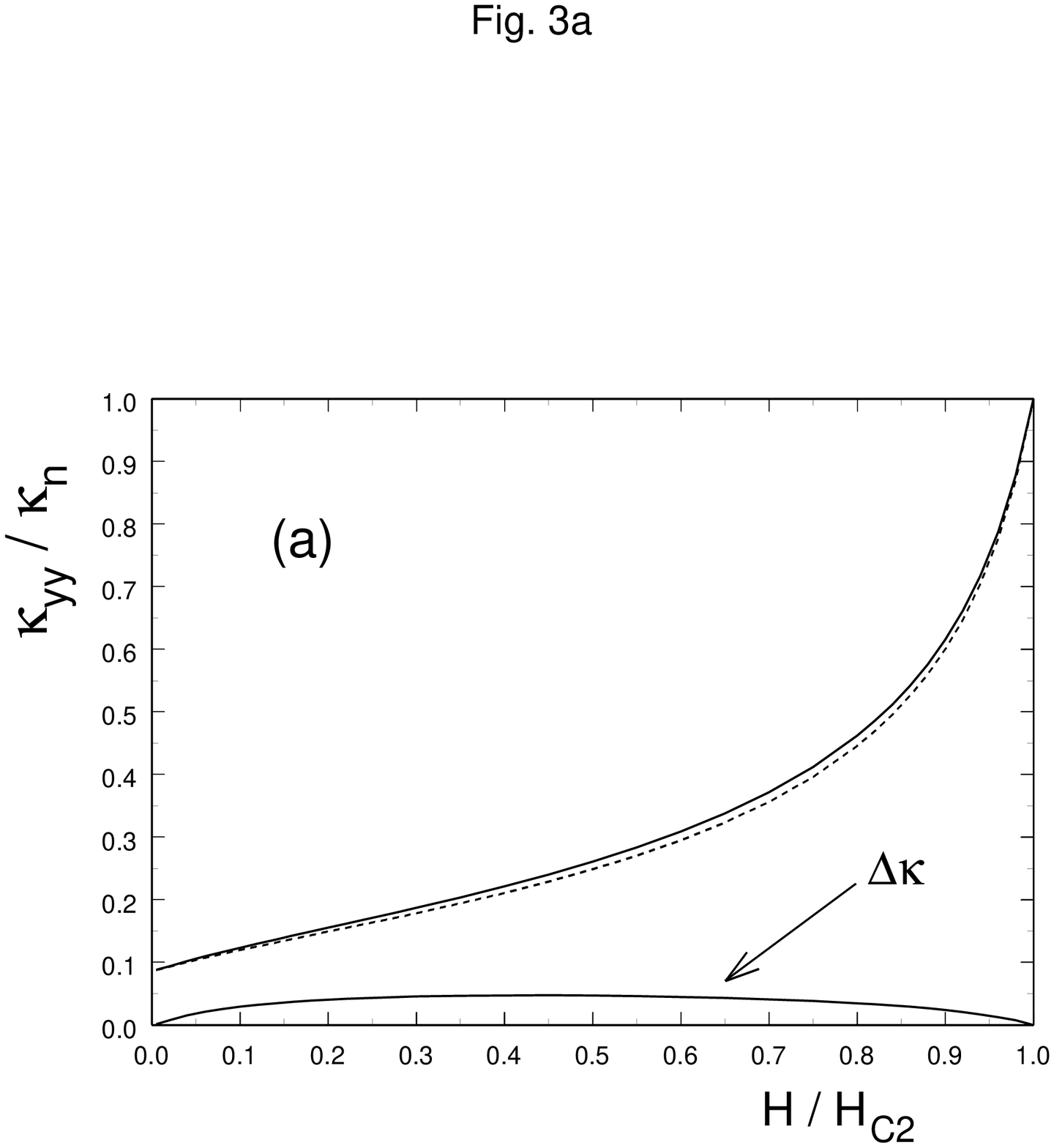,width=18cm,angle=0}}
\vskip -1cm
\caption{3a) $\kappa_{yy}$ and $\Delta\kappa$ vs $h\,$ (see notations in Fig. 1a) for
impurity scattering rate $\delta=0.1\,$, phase shift $\pi/2\,$, and finite
frequency (temperature) $\Omega=0.3\,$.}
\label{fig3a}
\end{figure}
\begin{figure}
\centerline{\psfig{file=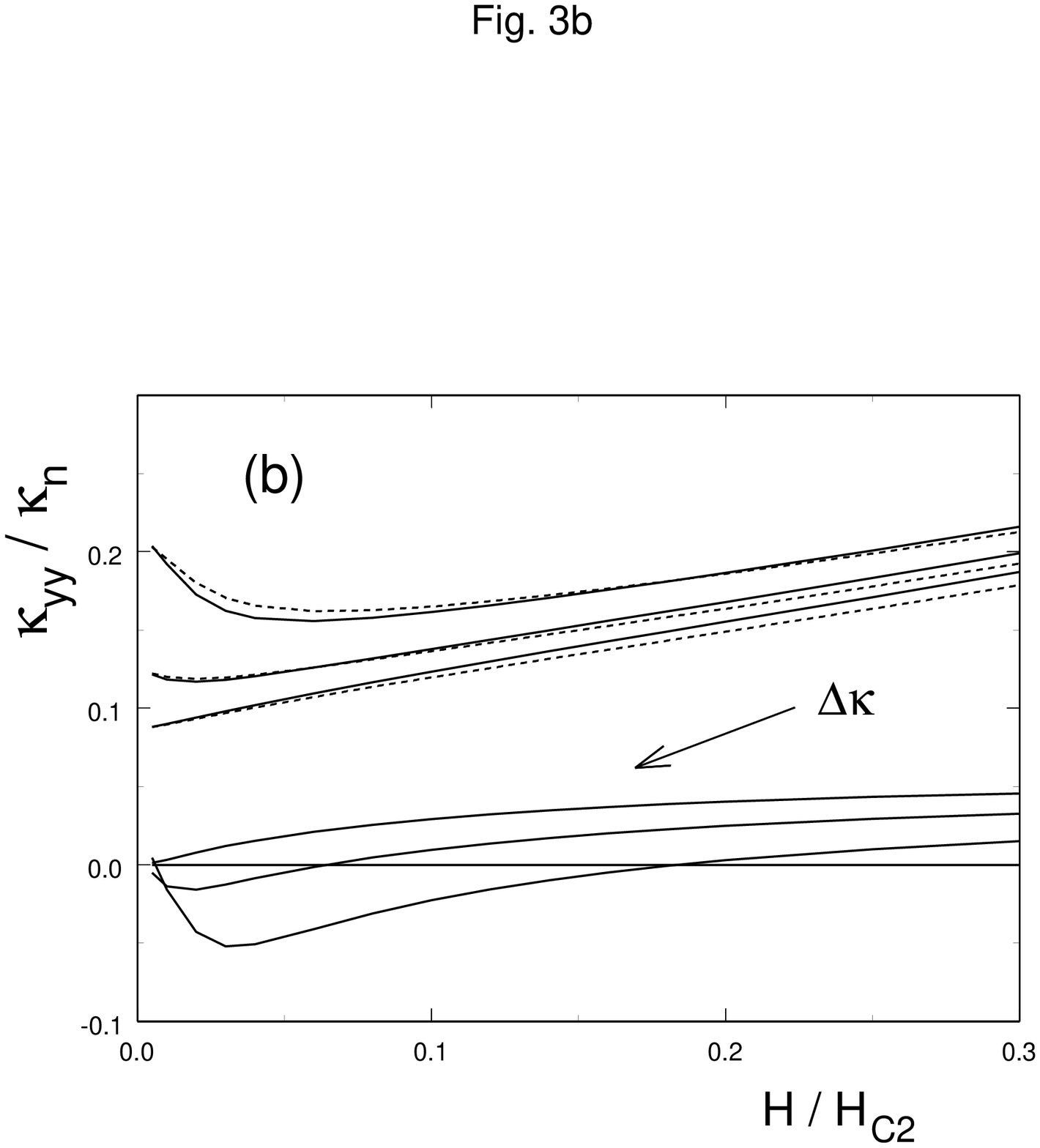,width=18cm,angle=0}}
\vskip -1cm
\caption{3b) $\kappa_{yy}$ vs $h$ for 
$\Omega=$ 0.3, 0.4 and 0.5 (upper curves from bottom to top) and 
$\Delta\kappa$ vs $h$ for $\Omega=$ 0.3, 0.4 and 0.5 
(lower curves from top to bottom).}
\label{fig3b}
\end{figure}

\begin{references}
%
%
\bibitem{Watanabe}T. Watanabe, K. Izawa, Y. Kasahara, Y. Haga. Y. Onuki,
P. Thalmeier, K. Maki, and Y. Matsuda, Phys. Rev. B {\bf 70}, 
184502 (2004).
%
\bibitem{McHale}P. McHale, P. Fulde, and P. Thalmeier,
Phys. Rev. B {\bf 70}, 014513 (2004).
%
\bibitem{Won}H. Won, D. Parker, K. Maki, T. Watanabe, K. Izawa, and
Y. Matsuda, Phys. Rev. B {\bf 70}, 140509(R) (2004).
%
\bibitem{Thalmeier}P. Thalmeier, T. Watanabe, K. Izawa, and Y. Matsuda, 
cond-mat/0501103.
%
\bibitem{BPT}U. Brandt, W. Pesch, and L. Tewordt, 
Z. Phys. {\bf 201}, 209 (1967).
%
\bibitem{TF1}L. Tewordt and D. Fay, Phys. Rev. B {\bf 64}, 024528 (2001).
%
\bibitem{Pesch}W. Pesch, Z. Phys. B {\bf 21}, 263 (1975); P. Klimesch and
W. Pesch, J. Low Temp. Phys. {\bf 32}, 869 (1978).
%
\bibitem{Vek}I. Vekhter and A. Houghton,  Phys. Rev. Lett.  {\bf 83}, 
4626 (1999).
%
\bibitem{TF2}L. Tewordt and D. Fay, cond-mat/0404610.
%
\bibitem{Dahm}T. Dahm, S. Graser, C. Iniotakis, and N. Schopohl, 
Phys. Rev. B {\bf 66}, 144515 (2002).
%
\bibitem{Nakai}N. Nakai, P. Miranovic, M. Ichioka, and K. Machida, 
Phys. Rev. B {\bf 70}, 100503 (R) (2004).
%
\bibitem{Graser}S. Graser, T. Dahm, and N. Schopohl, 
Phys. Rev. B {\bf 69}, 014511 (2004).
%
\end{references}
\end{document}